\documentclass[12pt]{article}

\usepackage{euscript}
\usepackage{amssymb}
\usepackage{amsfonts}
\usepackage{amsbsy}
\usepackage{amsmath}
\usepackage{epsfig}
\usepackage{amsthm}
\usepackage{amscd}
\usepackage{amstext}
\usepackage{verbatim}
\usepackage{textcase}
\usepackage[small,labelfont=bf,margin=1cm]{caption}

\usepackage[pdftex]{hyperref}
\usepackage{setspace}

\usepackage{feynmp}
\usepackage[explicit]{titlesec}
\usepackage[normalem]{ulem}

\titleformat{\section}
  {\normalfont\large\bfseries\filcenter}{\thesection.\hspace*{ 1em}}{0em}{#1}
  
\titleformat{\subsection}
  {\normalfont\it\filcenter}{\thesubsection\hspace*{1em}}{0em}{#1}

\renewcommand*{\thesection}{\arabic{section}}
\renewcommand*{\thesubsection}{\thesection.\arabic{subsection}}

 \numberwithin{equation}{section}
 
\usepackage{tocloft}

\cftsetindents{section}{0em}{2.5em}
\cftsetindents{subsection}{2.5em}{3.5em}
\def\ben{\begin{equation}}
\def\een{\end{equation}}
\def\half{{\textstyle{1\over2}}}

    \let\p=\phi

\def\be{\begin{equation}}
\def\ee{\end{equation}}
\def\beq{\begin{equation}}
\def\eeq{\end{equation}}
\def\ba{\begin{array}}
\def\ea{\end{array}}

\def\dalemb#1#2{{\vbox{\hrule height .#2pt
       \hbox{\vrule width.#2pt height#1pt \kern#1pt
               \vrule width.#2pt}
       \hrule height.#2pt}}}

\newcommand{\bea}{\begin{eqnarray}}
\newcommand{\eea}{\end{eqnarray}}

\newcommand{\Tr}{{\rm Tr} }

\renewcommand{\p}{\partial}

\newcommand{\bO}{\mathcal{O}}
\newcommand{\bh}{\bar{h}}
\newcommand{\bF}{\mathcal{F}}
\newcommand{\hn}{H}
\newcommand{\bz}{\bar{z}}
\newcommand{\bx}{\bar{x}}

\begin{document}

\begin{center}

{ \LARGE {\bf Entanglement Entropy \\ \bigskip at Large Central Charge}}
\vspace{1.2cm}

Thomas Hartman

\vspace{0.9cm}

{\it Institute for Advanced Study, School of Natural Sciences  \\
Princeton, NJ, USA
}

\vspace{0.5cm}

{\tt hartman@ias.edu} \\

\vspace{1.6cm}

\end{center}

\begin{abstract}
Two-dimensional conformal field theories with a large central charge and a small number of low-dimension operators are studied using the conformal block expansion.  A universal formula is derived for the Renyi entropies of $N$ disjoint intervals in the ground state, valid to all orders in a series expansion.  This is possible because the full perturbative answer in this regime comes from the exchange of the stress tensor and other descendants of the vacuum state. Therefore, the Renyi entropy is related to the Virasoro vacuum block at large central charge.  The entanglement entropy, computed from the Renyi entropy by an analytic continuation, decouples into a sum of single-interval entanglements.  This field theory result agrees with the Ryu-Takayanagi formula for the holographic entanglement entropy of a 2d CFT, applied to any number of intervals, and thus can be interpreted as a microscopic calculation of the area of minimal surfaces in 3d gravity.  

\end{abstract}

\pagebreak
\setcounter{page}{1}

\tableofcontents

\section{Introduction}

Entanglement entropy measures the entropy of a subsystem after tracing out the environment.  In two spacetime dimensions at a critical point, when the subsystem is a single interval, the ground state entanglement entropy is universal in the sense that it does not depend on the details of the conformal field theory \cite{Holzhey:1994we,Calabrese:2004eu,Calabrese:2009qy}. Conformal invariance fixes $S$ in terms of the central charge $c$ and the length $L$ of the subsystem.

The entanglement entropy of a subsystem that consists of multiple disjoint intervals depends, in general, on the full operator content of the theory \cite{Calabrese:2009ez,Calabrese:2010he}.  One aim of this paper is to show that in the limit of large central charge, in a theory with a small number of light operators, the entanglement entropy is again universal.  To leading order in $1/c$, it is uniquely fixed by conformal invariance, though in a more elaborate way than the entanglement of a single interval.  Renyi entropies, defined below, are similarly universal.

The motivation for studying this class of CFTs comes from the AdS/CFT correspondence, since these are the CFTs which are expected to have a holographic dual with a good semiclassical gravity description. Entanglement entropy is typically difficult to compute in field theory, but in a CFT with a holographic dual, there is a simple and universal proposal \cite{Ryu:2006bv}. For a single interval in two dimensions, it is straightforward to check explicitly that the holographic formula agrees with the standard CFT result. In higher dimensions or with multiple intervals, the simple formula is a surprise, but it has passed a number of nontrivial tests \cite{Takayanagi:2012kg}. A partial derivation exists for two intervals in two dimensions, for the first few orders in a series expansion \cite{Headrick:2010zt}. We extend this derivation to all orders and to any number of intervals. 

The result also applies to problems related by a conformal mapping, including the (possibly time dependent) entanglement entropy of a CFT at finite temperature where it is very natural to consider multiple intervals \cite{Azeyanagi:2007bj,Morrison:2012iz,Hartman:2013qma}.

The argument relies on a formula for the Virasoro conformal block at large $c$.  This formula is well known, but appears to have found few (if any) direct applications in AdS/CFT.  It is likely that other universal features of 3d gravity, or of 2d CFTs with gravity duals, can be understood in a similar way.  

The strategy to compute the entanglement entropy and the relevance of the conformal block are as follows. Divide a system, always taken to be in its groundstate, into two parts $A$ and $B$. The reduced density matrix of region $A$ is obtained by tracing out $B$, $\rho_A = \Tr_B \rho_{tot}$. This creates a mixed state, with entanglement entropy
\be
S_A = -\Tr \rho_A \log \rho_A \ .
\ee
One approach to compute this is the replica method. We define the Renyi entropies
\be
S_A^{(n)} = \frac{1}{1-n}\log \Tr \rho_A^n  \ ,
\ee
for integer $n\geq 2$, then analytically continue $n \to 1$ to find the entanglement entropy $S_A = S_A^{(1)}$. When $A$ consists of $N$ disjoint intervals, the Renyi entropy can be realized as a $2N$-point correlation function.  We will expand this correlation function in conformal blocks, and show that the leading contribution to the Renyi entropy at large $c$ is captured entirely by the Virasoro block for the vacuum state. Other contributions are exponentially suppressed. This statement is true to all orders in the OPE series expansion, but fails non-perturbatively as different terms in the expansion exchange dominance at large $c$. Note that unlike higher dimensions, the vacuum block in 2d CFT is nontrivial, since it includes the stress tensor and an infinite number of other Virasoro descendants.

The vacuum block is not known in closed form but can be computed numerically by a simple recursion relation to find the Renyi entropy to high accuracy.  It can also be computed by solving a certain monodromy problem for a second order differential equation.  In the limit $n \to 1$, the monodromy problem is solved analytically to compute the entanglement entropy.  The result for $N$ disjoint intervals is
\be\label{introres}
S_A = \frac{c}{3} \sum_{(i,j)} \log \left(z_i - z_j\over \epsilon\right)
\ee
where $z_i$ for $i=1\dots 2N$ is the endpoint of an interval, $\epsilon$ is a UV cutoff, and the sum is over pairs $(i,j)$ dictated by the OPE channel in a way described below.  This is equivalent to the holographic formula \cite{Ryu:2006bv}.  

The CFT derivation of (\ref{introres}) is valid only within some finite region around the origin of the OPE in any channel. A complete derivation of the holographic formula would require a non-perturbative argument that there are no other phases in the parameter space of $z_i$, i.e., phases which do not include the origin of any OPE channel. Of course, the full OPE at finite central charge is convergent and any $z_i$ can be expanded in some channel; however, the leading-$c$ term is not analytic in the same range of $z_i$ as the original OPE. We show that crossing symmetry rules out any other phases for $N=n=2$ but do not address the general case non-perturbatively.

The Renyi entropies also match precisely with a gravity calculation done in \cite{faulkner}.  In fact, exactly the same monodromy prescription for computing $S_A^{(n)}$ that we will derive from CFT was derived independently from 3d gravity in a completely different way \cite{faulkner}.  This means that the semiclassical conformal block with external twist operators and an internal unit operator is equal to the on-shell Einstein action on an appropriate 3-manifold with nontrivial topology.

As mentioned above, the technique used to derive the Renyi entropy may have other applications in 3d gravity, for example to the calculation of black hole scattering amplitudes from CFT. It is a manifestation of the known connection between AdS$_3$ gravity, $SL(2,R)$ Chern-Simons theory, and Liouville CFT in the classical limit \cite{Achucarro:1987vz,Witten:1988hc,Verlinde:1989ua,Coussaert:1995zp}.  We will not explore this triangle of connections in any detail, but return to this point of view in the discussion section.

\section{Semiclassical conformal blocks}

We begin with a discussion of correlation functions in a 2d CFT with large central charge using the OPE.  The main conclusion will be that in this limit, the stress tensor and other Virasoro descendants of the vacuum give a nontrivial contribution to the 4-point function that can be computed by imposing a trivial monodromy condition on a certain differential equation. 

\subsection{General operators}\label{genop}
A four-point function of primary operators on the plane may be expanded in conformal blocks,
\be\label{ope}
\langle \bO_1(0)\bO_2(x)\bO_3(1)\bO_4(\infty) \rangle = \sum_p a_p \bF(c, h_p, h_i, x)\bF(c,\bh_p, \bh_i, \bar{x}) \ .
\ee
We have expanded in the $s$-channel $x\to 0$. Here $(h_i, \bh_i)$ are the dimensions of $\bO_{i=1\dots 4}$, the sum is over primary operators $\bO_p$ of dimension $(h_p, \bh_p)$, and 
\be
a_p = c^{p}_{12}c^p_{34}
\ee
where $c^p_{ij}$ is an OPE coefficient.  The Virasoro blocks $\bF$ capture the contribution of all the Virasoro descendants. They are unknown in general, but straightforward to compute in a series expansion. This can be done very efficiently using a recursion formula \cite{zam1} described in appendix \ref{app:numerics}.

We are interested in correlation functions at large central charge.  In the `semiclassical' limit, defined by taking $c$ large with $h_i/c$ and $h_p/c$ held fixed, the block exponentiates \cite{Belavin:1984vu,zam1},
\be\label{expo}
\mathcal{F}(c,h_p, h_i, x) \approx \exp\left[-\frac{c}{6} f\left(\frac{h_p}{c}, \frac{h_i}{c}, x\right)\right] \ .
\ee
The function $f$ is once again unknown, except in an expansion around $x=0$:
\be\label{fseries}
\frac{c}{6}f  = (h_1+h_2-h_p)\log x -  \frac{(h_p+h_2 -h_1)(h_p + h_3-h_4)}{2h_p}x + O(x^2) \ .
\ee
However $f$ is determined by the solution of a certain monodromy problem.  Consider the differential equation
\be\label{diff1}
\psi''(z) + T(z) \psi(z) = 0 \ ,
\ee
with 
\be\label{diff2}
T(z) = \sum_{i} \left( \frac{6 h_i/c}{(z-z_i)^2} - \frac{c_i}{z-z_i}\right) \ ,
\ee
where $(z_1, z_2, z_3, z_4) = (0,x,1,\infty)$.  The $c_i$ are called accessory parameters. Three of them are fixed by the requiring $T(z)$ to vanish as $z^{-4}$ at infinity. This imposes
\be\label{treg}
\sum_i c_i  = 0  \ , \quad \sum_i(c_i z_i - \frac{6 h_i}{c}) = 0 \ , \quad \sum_i(c_i z_i^2 - \frac{12 h_i}{c}z_i) = 0 \ ,
\ee 
so that
\be
\frac{c}{6}T = \frac{ h_1}{z^2} + \frac{h_2}{(z-x)^2} + \frac{h_3}{(1-z)^2} + \frac{h_1+h_2+h_3 -h_4}{z(1-z)} -\frac{c}{6} \frac{c_2 x(1-x)}{z(z-x)(1-z)} \ .
\ee
The differential equation (\ref{diff1}) has two solutions, $\psi_1$ and $\psi_2$. As we take the solutions on a closed contour around one or more singular points, they undergo some monodromy
\be\label{monoto}
{\psi_1 \choose \psi_2}  \rightarrow M {\psi_1 \choose \psi_2} \ .
\ee
The 2x2 monodromy matrix $M$ depends on the basis of solutions $\psi_{1,2}$, but its trace is invariant. 

The connection between this differential equation and the semiclassical conformal block is the following.  First, we choose the accessory parameter $c_2(x)$ so that the monodromy on a cycle enclosing both 0 and $x$ is 
\be\label{genmono}
\Tr M_{0x} = -2\cos\pi \Lambda_p \ ,\quad h_p = \frac{c}{24}(1-\Lambda_p^2) \ .
\ee
Then the semiclassical Virasoro block is determined by
\be
\frac{\p f}{\p x} = c_2(x) \ .
\ee
The integration constant in this equation is fixed by comparing to the series expansion (\ref{fseries}).

Although the formula (\ref{expo}) has never been proved directly from the definition of the Virasoro block as a sum over descendants, it follows from the path integral of Liouville theory, and has passed extensive checks. We will review the Liouville derivation in section \ref{ss:blockn} below and generalize it to more than four external operators.

The Virasoro conformal block depends only on the algebra, so this result from Liouville theory is applicable to any CFT in the semiclassical limit.

\subsection{Unit operator}\label{ss:unit}

The semiclassical limit $c \to \infty$ is usually taken with $h_p/c$ held fixed.  In fact, if the external weights are equal, $h_i = h$, then we can also take the semiclassical limit with $h_p$ held fixed, i.e. $\gamma\equiv h_p/c \to 0$. The limits commute:
\be\label{comlim}
\lim_{\gamma \to 0} \lim_{c\to \infty} \frac{1}{c}\log \bF(c, c \gamma, c \beta, x) = \lim_{c\to\infty}\frac{1}{c}\log \bF(c, 0, c\beta, x) \ .
\ee
This can be seen from the recursion representation of $\bF$ in appendix \ref{app:numerics}.\footnote{In the recursion, the series expansion of $\log \bF$ is organized so that the only $h_p$ dependence is a sum over terms of the form $\frac{a}{(h_p - b)^k}$ with $b = O(c)$. The limits in (\ref{comlim}) commute term by term. It is enough for the external weights to be equal in pairs, $h_1 = h_2$, $h_3 = h_4$.} Let us denote the semiclassical block for these light operators by
\be
f_0\left(\frac{h}{c}, x\right) = f\left(0, \frac{h}{c}, x\right) \ .
\ee
The dependence on $h_p$ disappears, so we will set $h_p=0$ and refer to this as the vacuum block. In $d>2$ spacetime dimensions, the vacuum block is trivial; it gives only the disconnected piece of a correlator.  However in $d=2$, the vacuum block includes the stress tensor and all of the other Virasoro descendants.

It can be computed (in principle, or numerically) by solving the monodromy problem above, where now we tune $c_2(x)$ to impose trivial monodromy around the points $0,x$:
\be\label{trivm}
\Tr M_{0x} = 2  \ .
\ee
This follows from (\ref{genmono}) with $h_p = 0$. The semiclassical vacuum block can also be interpreted as a classical on-shell Liouville action; we will return to this below.

\section{Entanglement entropy of two intervals}\label{s:twointervals}

\subsection{CFT calculation}

Suppose that we have a family of theories, labeled by the central charge, that admits a large-$c$ limit. Besides unitarity and compactness, we will make only two assumptions about this family of CFTs.  First, in the large-$c$ limit, correlation functions are smooth in a neighborhood of coincident points and obey cluster decomposition.  This restricts the growth of OPE coefficients to be at most exponential in $c$; otherwise the vacuum does not appear in the OPE as $x \to 0$.  Second, we assume that the number of light operators does not grow with $c$. Specifically, the density of states is $d(\Delta,\bar{\Delta}) = O(c^0)$ for $\Delta,\bar\Delta < \Delta_{gap}$ where we will take $\Delta_{gap} = c/24$ but this can be relaxed to $\Delta_{gap} = O(c)$. Examples are symmetric product theories and theories with a gravity dual (in this case $\Delta_{gap}$ is the energy of the lightest black hole). We will compute the Renyi entropies and entanglement entropy in theories of this type to leading order in $1/c$, without using the specific details of the CFT.

The replica method, described in the introduction, is a useful approach to entanglement entropy when the system is in a state where the wavefunction is computed by a path integral on some manifold $M$. This includes the vacuum state, or a thermal state in CFT.  In this case the replica partition function can be characterized in two equivalent ways.  First, it is a path integral on an $n$-sheeted cover of $M$, with branch cuts on region $A$ where the sheets are connected. This manifold is a singular Riemann surface with nontrivial topology.  Second, the partition function can be computed in an $n$-fold product theory on the original manifold $M$, with twist fields inserted at the boundaries of region $A$. The twist fields are defined to glue together the $n$ copies of the CFT in a way that reproduces the path integral on the higher genus surface (see \cite{Calabrese:2009qy} for a review).\footnote{It is not guaranteed that the replica trick will always produce the correct entanglement entropy because of ambiguities in the analytic continuation $n \to 1$.  We will assume without proof that it works with the obvious choice of analytic continuation in the CFTs we consider, and furthermore that we can perform the analytic continuation on the leading-$c$ term directly, but we caution that this is not entirely justified.}

For region $A$ consisting of $N$ disconnected intervals,
\be\label{amany}
A = [z_1,z_2] \cup [z_3,z_4] \cup \cdots \cup [z_{2N-1},z_N] \ , \quad\quad z_1 < z_2 < \cdots < z_{2N}  \ ,
\ee
the Renyi entropy, computed in the product theory on $M$, is \cite{Calabrese:2004eu,Calabrese:2009qy}
\be\label{gent}
\exp\left((1-n)S_A^{(n)}\right) = \langle \Phi_+(z_1)\Phi_-(z_2) \cdots \Phi_+(z_{2N-1})\Phi_-(z_{2N})\rangle \ .
\ee
The twist operators $\Phi_\pm$ have weight
\be\label{tdim}
(L_0, \bar{L}_0) = (\hn, \hn) \  , \quad\quad \hn = \frac{c}{24}(n-1/n) \ .
\ee
For a single interval, $S_A^{(n)}$ is completely fixed by conformal invariance, but for $N\geq 2$ it generically depends on the full operator content of the theory.

In the rest of this section we set $N=2$.  This case has been discussed in detail in \cite{Headrick:2010zt}. We will repeat parts of that discussion in our language, both in order to setup the problem of general $N$ and to clarify and extend some aspects of the CFT calculation in \cite{Headrick:2010zt}.

We can set $(z_1,z_2, z_3, z_4) = (0,x,1,\infty)$ by a conformal transformation. The cross-ratio $x$ is real. The Renyi entropy is a 4-point function, so it has the conformal block expansion (\ref{ope}). At large central charge, expanding in the $s$-channel,
\be\label{twistope}
\exp\left((1-n)S_A^{(n)}\right) = \sum_p a_p \exp\left[-\frac{nc}{6}f\left(\frac{h_p}{nc},\frac{\hn}{nc}, x\right) - \frac{nc}{6}f\left(\frac{\bh_p}{nc}, \frac{\hn}{nc}, \bx\right)\right] \ ,
\ee
where we have used the central charge of the replica theory, $nc$. We want to show that the exponential dependence is entirely captured by the first term in the sum; low-lying terms affect only the irrelevant coefficient of the exponential, and high-dimension terms are non-perturbative in $1/c$.  This is typical behavior of a thermodynamic function with different phases.  The argument is simplest if we assume $a_p$ does not grow exponentially with $c$.\footnote{
This extra assumption is not necessary, though without it the region in which the phase (\ref{sn}) dominates may cover a smaller range of $x$.  The assumption that we have smooth correlators in a neighborhood of $x=0$ as $c\to \infty$ already implies a bound of the form
\be
a_p \leq \exp\left[\frac{c}{6}g\left(\frac{h_p}{nc}, \frac{\bar{h}_p}{nc}\right)\right] \ .
\ee
Therefore, up to an irrelevant prefactor to account for multiplicities, we can bound the sum over low-dimension operators in (\ref{twistope}) by
\be
\int_0^{1/24} d\delta_p \int_0^{1/24} d\bar{\delta}_p \exp\left[\frac{c}{6}\left(g(\delta_p, \bar\delta_p) - f(\delta_p, \frac{H}{nc},x) -f(\bar\delta_p, \frac{H}{nc}, \bar{x})\right)\right] \ .
\ee
This is exponentially dominated by an endpoint of the limits of integration, or by a saddlepoint; near $x=0$, it must be dominated by $\delta_p = \bar{\delta}_p = 0$, because the vacuum must give the leading term in the correlation function near coincident points. }  Then our assumption about the low-lying operator spectrum allows us to ignore the coefficient and any multiplicities in (\ref{twistope}) for $h_p,\bar{h}_p < \Delta_{gap}$.  In this range, $f(\delta_p, \frac{H}{nc}, x)$ is an increasing function of $\delta_p$ for any $x < \half$, so the low-lying terms exponentially dominate.  Similarly, heavy operators in (\ref{twistope}) are suppressed non-perturbatively in $1/c$.  Therefore in a neighborhood of $x=0$
\be\label{sn}
S_A^{(n)}  \approx \frac{nc}{3(n-1)} f_0\left(\frac{H}{nc}, x\right) \ .
\ee
This is one of our main results.  It is the contribution from all light operators, or equivalently the contribution from the vacuum and its descendants. We reiterate that (\ref{sn}) is the full answer for the Renyi entropy to leading order in $1/c$ in a \textit{finite} region around $x = 0$, i.e., to all orders in a series expansion in $x$. We will return to the question of how big this region is below. There are both perturbative and non-perturbative corrections in $1/c$. We have not used any special properties of the twist operators other than the fact that $H \sim c/24$, so a similar formula applies to the leading-$c$ correlation function of other heavy operators. 

The result (\ref{sn}) allows for a simple numerical or series calculation of the Renyi entropy for any $n$, not necessarily an integer.  This can be done easily using the recursion formula in appendix \ref{app:numerics} to compute $\log \bF$. The results for  various $n$ are plotted in figure \ref{fig:renyiplots}(a), and the series expansion is given in the appendix. Alternatively, it can be computed by solving the trivial-monodromy problem that defines $f_0$, but for general $n$ this is more difficult than just computing $\log \bF$ directly.  

\begin{figure}
\begin{center}
\includegraphics{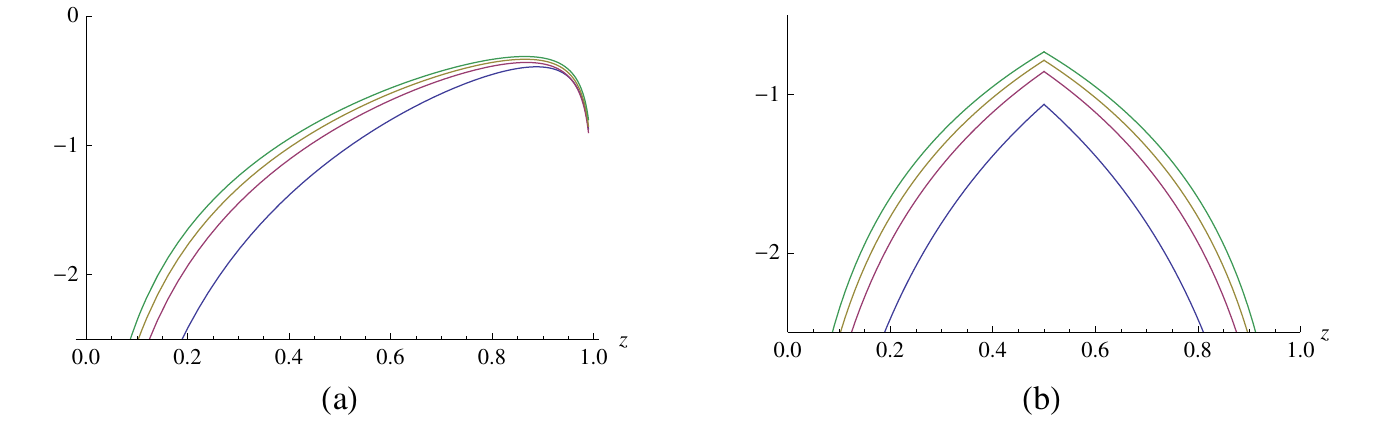}
\end{center}
\caption{\footnotesize The vacuum contribution to the Renyi entropy (divided by $c/6$), for $n=2,5,10,40$, from bottom to top. In (a) we show the contribution to the $s$-channel, and in (b) both channels.  These are computed from (\ref{sn}) using the recursion relation in appendix \ref{app:numerics} iterated 10 times. The error, estimated by adjusting $c$ and discarding the last term in the recursion, is less than 1\% for $z<0.99$. The $\log \epsilon$ UV divergence is dropped.
\label{fig:renyiplots}}
\end{figure}

The advantage of the monodromy prescription is that it can be solved analytically in the limit $n \to 1, H\to 0$ relevant for the entanglement entropy.  Denoting $n = 1 + 12 \alpha$, the entanglement entropy is
\be\label{enta}
S_A =  \frac{c}{36} \lim_{\alpha \to 0}\frac{1}{\alpha}f_0(\alpha,x) \ ,
\ee
so we only need to solve the monodromy problem with vanishingly small operator insertions.  For small $h_i$ in (\ref{diff2}), the accessory parameter $c_2$ will also be small, so all terms in $T(z)$ are important only at the singular points.  Therefore to solve the monodromy problem around $0,x$, we can ignore the other singular points, and consider the simpler problem with
\be\label{tsimp}
T(z) = 6\alpha\left( \frac{1}{z^2} + \frac{1}{(z-x)^2} + \frac{2}{z} \right) + \frac{c_2(1-x)}{z} - \frac{c_2}{z-x} \ .
\ee
This has trivial monodromy around $0,x$ if $T$ is regular at infinity. Therefore we require the sum of the residues of simple poles to vanish:
\be
c_2 = \frac{12 \alpha}{x} + O(\alpha^2) \ .
\ee
Integrating, 
\be\label{firstf}
f_0(\alpha,x) = 12 \alpha \log x  + O(\alpha^2)\ .
\ee
This is the leading-$c$ contribution to the Virasoro block for a heavy internal operator, analytically continued to small $h_p$; note that it is not equal to the block for a light internal operator. The formula (\ref{firstf}) states that only the identity operator -- not the stress tensor or its descendants -- contributes, at any value of $x$.   Thus the entanglement entropy for two intervals will factorize, although this was not true of the Renyi entropies with $n>1$. From (\ref{enta}) we find the entanglement entropy
\be\label{sanswer}
S_A =  \frac{c}{3} \log \left(\frac{x}{\epsilon}\right) \quad \quad (s\mbox{-channel}) \ .
\ee
(In this expression and similar expressions below we have reintroduced the UV cutoff $\epsilon$ that is necessary to regulate the twist operators.)

We chose to expand in the $s$-channel $x \sim 0$, but now let us expand in the $t$-channel $x \sim 1$.  Blocks in the $t$-channel are related to those in the $s$-channel by $x \to 1-x$ so the answer can be obtained from (\ref{sanswer}) but let us demonstrate this directly. The argument is identical, except that now we define $f_0$ by imposing trivial monodromy around a cycle enclosing $x,1$ instead of $0,x$. Instead of (\ref{tsimp}) we have
\be
T(z) = 6\alpha\left( \frac{1}{(z-1)^2} + \frac{1}{(z-x)^2} - \frac{2}{z-1}\right) + \frac{c_2 x}{z-1} - \frac{c_2}{z-x} \ .
\ee
Regularity implies $c_2 = 12\alpha/(x-1)$, and $\p f_0/\p x = c_2$ gives
\be
f_0 = 12 \alpha \log (1-x) +O(\alpha^2)\ .
\ee
Therefore
\be\label{tent}
S_A = \frac{c}{3}\log \left(\frac{1-x}{\epsilon}\right)\quad \quad (t\mbox{-channel}) \ ,
\ee
to all orders in a series expansion around  $x=1$.

The results (\ref{sanswer}, \ref{tent}) were derived to order $x^5$ by a series expansion of the Virasoro blocks in \cite{Headrick:2010zt}.  From the present derivation, which follows similar logic but exploits the formula for the semiclassical blocks, it is valid in finite regions around $x=0$ and $x=1$.

However we have not specified the range of $x$ in which (\ref{sn}, \ref{sanswer}, \ref{tent}) are valid.  To do this we would need to determine when the heavy operators in (\ref{twistope}) first start to dominate.  The $s$-channel result (\ref{sanswer}) must break down at or before $x=\half$, because at this point, the heavy operators in the $s$-channel must account for the vacuum running in the $t$-channel.  The question is whether there is another transition at some point $x=x_c$, with $0 < x_c < \half$, to a phase that does not correspond to the vacuum block in either channel. (Other phases certainly exist elsewhere on the complex plane, but in our case $x$ is real and $0<x<1$.) Such a phase would not be visible in a perturbative expansion (in $1/c$) around the origin of the OPE.

The answer for $n=2$ is that the $s$-channel Renyi entropy is valid for $0 < x < \half$, and there is a sharp transition at $x=\half$ to the $t$-channel result. There are no other phases. This was demonstrated in \cite{Headrick:2010zt} by mapping the $n=2$ twist correlator to a torus partition function.  An alternate derivation is given in appendix \ref{app:torus} using crossing symmetry of the OPE to show that the contribution of heavy operators in the $s$-channel is equal to the vacuum term in the $t$-channel. The argument is limited to $n = 2$ unless we add the assumption that certain OPE coefficients are exponentially suppressed at large $c$, so we leave open the question of whether there are other phases in the higher Renyi entropies or the entanglement entropy. It may depend on the particular theory.  If we assume that the dominant contribution is always the vacuum in some channel, then the entanglement entropy is
\be
S_A = \frac{c}{3}\min\left\{ \log\left(\frac{x}{\epsilon}\right) \ , \ \log\left(\frac{1-x}{\epsilon}\right) \right\} \ ,
\ee
and the Renyi entropies for various $n$ are as shown in figure \ref{fig:renyiplots}(b). 

\subsection{Comparison to holography}\label{ss:compare}

In \cite{Ryu:2006bv} it was conjectured that the entanglement entropy of a CFT with a holographic dual can be computed by a simple geometric formula.  The proposal is 
\be
S_A = \mbox{min}_\gamma \ \frac{\mbox{area}(\gamma)}{4 G_N} \ ,
\ee
where $\gamma$ is a surface of dimension $d-1$ in a fixed-time slice of AdS$_{d+1}$ that meets the boundary of region $A$ at the boundary of AdS. In AdS$_3$, Newton's constant is $G_N = \frac{3}{2c}$, and the `area' is the length of a spacelike geodesic. When $A$ is a disconnected region, the bulk surface can also be disconnected, and it is the total area that computes the entropy.

The result when $A$ is two disjoint intervals agrees with the CFT calculation above \cite{Headrick:2010zt}.  We have two choices for how to draw the geodesics that end on the endpoints of region $A$:
\begin{equation*}
\includegraphics{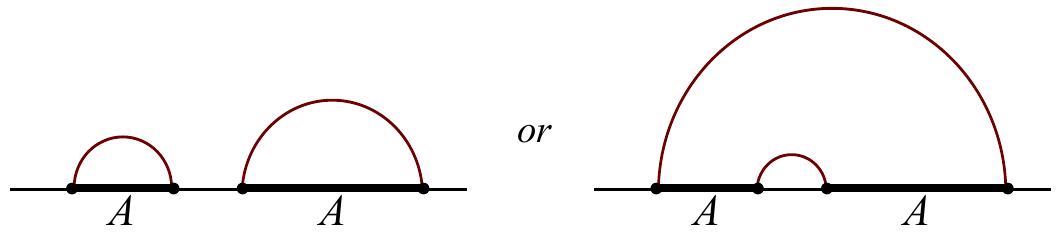}\quad\quad .
\end{equation*}
The configuration that gives the entropy is the one in which the geodesics have minimal total length.  This is the disconnected one for small cross-ratio $x$ and the connected one for large $x$, with a sharp transition at $x=\half$. Note that in any particular channel, the answer decouples into a sum of single-interval entanglements.  From the CFT point of view, this reflects the decoupling of the different monodromies that must be imposed to compute the semiclassical conformal block, when the singularities of the differential equation have vanishing weight as $n\to 1$.

The holographic entanglement entropy of a single interval gives the classic CFT result mentioned in the introduction, $S_{ij} = \frac{c}{3}\log\left(z_i - z_j\over\epsilon\right)$.  From the diagram, the holographic result for two intervals is
\be
S_A = \min\ (S_{12} + S_{34} \  , \ \  S_{14} + S_{23} ) .
\ee
This agrees with the CFT calculation, applied to the entire range $0 < x < 1$. Assuming the holographic formula is correct, this suggests that in these CFTs the leading contribution to the Renyi entropy indeed always comes from the vacuum block in some channel. A closely related conjecture about the classical Einstein action on manifolds with a higher-genus boundary was made in \cite{Yin:2007gv}.

The Renyi entropies $S_A^{(n)}$ given in (\ref{sn}) also agree with a holographic calculation, performed in \cite{faulkner}. The match provides a very simple geometrical interpretation of the semiclassical vacuum block $f_0$ with external twist operators: it is the Einstein action of a particular 3-manifold.  We will review the gravity calculation briefly. See \cite{faulkner} for details.

The Renyi entropy is the partition function on a singular Riemann surface $M_2$ of genus $n-1$. In AdS/CFT, this partition function is related to the on-shell Einstein action of a spacetime $M_3$ whose conformal boundary is $M_2$. In general, there are many manifolds $M_3$ with the same conformal boundary, and we must choose the dominant saddlepoint.  Within a finite region around the origin of any OPE channel, we may restrict our attention to a certain class of handlebodies: given a genus-$g$ Riemann surface, we choose $g$ independent cycles $\Gamma_{1,\dots,g}$ to `fill in' to construct the bulk spacetime.  The on-shell Einstein action on a 3-manifold $M_3$ constructed in this way is equal to an appropriately defined Liouville action  on $M_2$ \cite{Krasnov:2000zq,Krasnov:2001ui,Krasnov:2002rn,faulkner},
\be
e^{-S_{Einstein}(M_3)} = e^{-\frac{c}{3}S_{ZT}(M_2, \Gamma_i)} \ .
\ee
The Liouville action $S_{ZT}$, defined in \cite{ztschottky}, depends on $M_2$ and on the choice of contractible cycles.  It was demonstrated in \cite{faulkner}, using \cite{ztsphere,ztschottky,Takhtajan:2001uj}, that $S_{ZT}$ is the generating function of the accessory parameters for the trivial-monodromy problem defined in section \ref{ss:unit}, $\p S_{ZT}/\p z_i = nc_i$. This is exactly the same prescription derived from the conformal blocks, so $S_{ZT} = nf_0$.
It follows that the Renyi entropy computed from CFT agrees with the gravitational partition function, with classical Liouville providing a natural link between the two calculations.\footnote{The semiclassical block $f$ also appears in other contexts, including the Bethe ansatz and 4d gauge theory.  See \cite{Nekrasov:2011bc,Gaiotto:2011nm} and references therein.} The choice of trivial-monodromy cycles in the definition of $f_0$, i.e. the choice of OPE channel, corresponds to the choice of cycles on $M_2$ that are filled in to construct the spacetime $M_3$.

\section{More than two intervals}
In this section we generalize the discussion to the entanglement entropy of $N>2$ intervals. Our aim is to compute the $2N$-point twist correlator (\ref{gent}) at large $c$, to all orders in a series expansion around coincident points. This is conceptually very similar to the case $N=2$: we find a monodromy prescription for the large-$c$ conformal blocks, then analytically continue $n\to 1$ where the monodromy problem can be solved analytically. 

\subsection{Semiclassical $k$-point conformal blocks}\label{ss:blockn}

First we must show that any $k$-point conformal block in the semiclassical limit can be computed by solving a monodromy problem similar to the one in section \ref{genop}. We will derive this from the path integral of the Liouville CFT at large $c$, generalizing the argument for the 4-point function \cite{zam1} as reviewed in \cite{Hadasz:2005gk,Harlow:2011ny}. The blocks are determined solely by the Virasoro algebra so the result applies to any CFT at large $c$.

First we describe the final result, since this is the only fact needed for the rest of the paper. It is the obvious generalization of section \ref{genop}.  A general $k$-point function on the plane
\be
\langle O_1(0)O_2(z_2)O_3(z_3)\cdots O_{k-1}(1) O_k(\infty)\rangle
\ee
can be expanded in many different channels.  The channels are represented by tree graphs with propagators and 3-point vertices.
An example for the 6-point function is the channel
\vspace{.3cm}
\begin{equation}\label{sixc}\small
\begin{fmffile}{sixA}
\begin{fmfgraph*}(60,20)
  	\fmfleft{i1,i2}
        \fmfright{o1,o2}
	\fmftop{t1,t2,t3,t4,t5,t6}
	\fmfbottom{b1,b2,b3,b4,b5,b6}
        \fmf{plain}{i1,v1,i2}
        \fmf{plain}{o1,v2,o2}
        \fmf{plain,label=$O_p(0)$}{v1,vA}
	\fmf{plain,label=$O_q(0)$}{vA,vB}
	\fmf{plain,label=$O_r(1)$}{vB,v2}
	\fmf{plain,tension=0}{t3,vA}
	\fmf{plain,tension=0}{b4,vB}
	\fmflabel{1}{i1}
	\fmflabel{2}{i2}
	\fmflabel{3}{t3}
	\fmflabel{4}{o2}
	\fmflabel{5}{o1}
	\fmflabel{6}{b4}
\end{fmfgraph*}
\end{fmffile}
\end{equation}
where $p,q,r$ label primaries appearing in the OPEs. The conformal block in this channel is defined so that
\be
\langle O_1 O_2 O_3 O_4 O_5 O_6 \rangle = \sum_{p,q,r}c_{12}^pc_{p3}^qc_{45}^rc^r_{q6}\bF(z_i) \bF(\bar{z_i})\ .
\ee
More generally, we draw a tree diagram with $k$ external points marked $1\dots k$ in cyclic order, and label each internal line by a primary operator. The blocks can be computed in a series expansion by summing over Virasoro descendants up to a given level \cite{Sonoda:1988mf,Moore:1988qv}, as described for example in \cite{Alday:2009aq,Alba:2009ya}.

In the semiclassical limit $c\to\infty$ with $h_i/c$ held fixed, the $k$-point Virasoro block in any channel exponentiates,
\be\label{blockup}
\mathcal{F}  \approx \exp\left[-\frac{c}{6}f\left(\frac{h_a}{c}, \frac{h_i}{c}, z_i\right)\right]
\ee
where $h_i$ denote the external weights and $h_a$ the internal weights.  The semiclassical block $f$ can be computed as follows.  Consider the differential equation (\ref{diff1}) with $T(z)$ as in (\ref{diff2}), now summing over $i=1\dots k$. Regularity imposes three conditions (\ref{treg}), so there are $k-3$ accessory parameters, which can be used to tune the $k-3$ independent monodromies of the differential equation. Choose cycles around the singular points which correspond to the OPE contractions in the chosen channel. That is, for each contraction $O_A(z_A) O_B(z_B) \to O_C(z_A)$, we choose a contour $\gamma_C$ enclosing $z_A$ and $z_B$.  In the 6-point example above, the cycles can be chosen as follows:
\begin{equation}\label{cycb}
\includegraphics[width=140px]{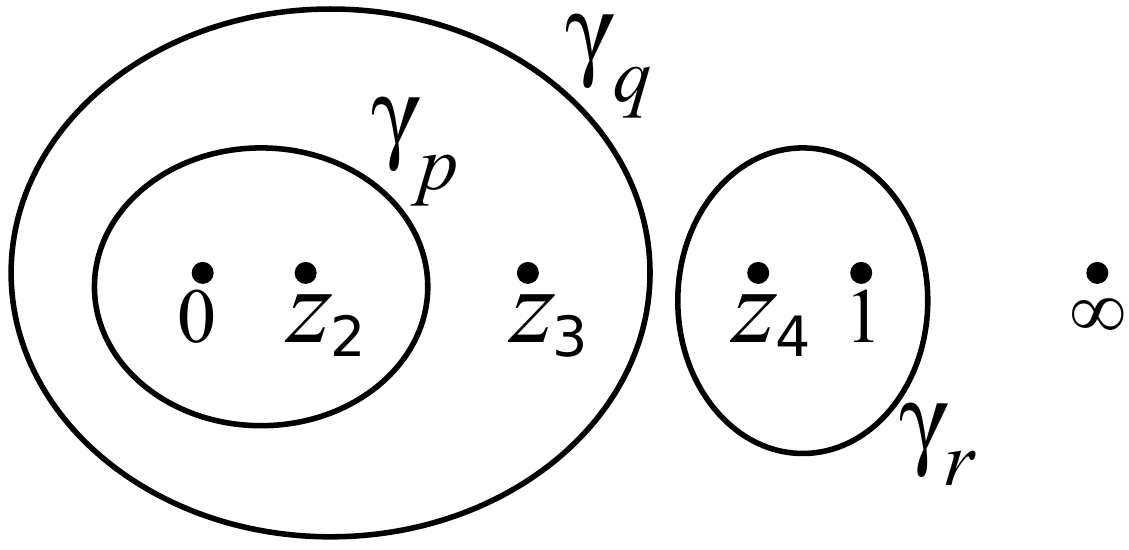}
\end{equation}
Now, choose the accessory parameters $c_{i=2,\dots,k-2}$ so that the monodromies on these cycles are
\be
\Tr M_a = -2 \cos \pi \Lambda_a \ , \quad h_a = \frac{c}{24}(1-\Lambda_a^2) \ ,
\ee
where $h_a$ is the weight of the primary operator appearing in the corresponding OPE channel. Finally, the semiclassical block is computed by integrating
\be
\frac{\p f}{\p z_i} = c_i \ .
\ee

We now restrict to equal external weights $h_i = h$. What is the analog of the `vacuum block' that was important in the previous discussion? This depends on the channel.  In some channels, we can choose every internal weight to vanish,
\be
f_0\left(\frac{h}{c}, z_i\right) = f\left(0,\frac{h}{c}, z_i\right)  \ .
\ee
This is possible only if the external operators are contracted in pairs, such as in the channels
\begin{equation*}\small
\begin{fmffile}{inpairs}
\parbox{20mm}{
\begin{fmfgraph*}(20,20)
	\fmfsurround{v1,v2,v3,v4,v5,v6}
	\fmf{plain}{v1,vA}
	\fmf{plain}{v2,vA}
	\fmf{plain}{v3,vB}
	\fmf{plain}{v4,vB}
	\fmf{plain}{v5,vC}
	\fmf{plain}{v6,vC}
	\fmf{plain}{vA,vX}
	\fmf{plain}{vB,vX}
	\fmf{plain}{vC,vX}
\end{fmfgraph*}}
\quad\quad \mbox{and} \quad\quad \parbox{20mm}{
\begin{fmfgraph*}(20,20)
	\fmftop{v1,v2,v3,v4}
	\fmfbottom{v5,v6,v7,v8}
	\fmf{plain}{v1,vA}
	\fmf{plain}{v2,vA}
	\fmf{plain}{v3,vB}
	\fmf{plain}{v4,vB}
	\fmf{plain}{v5,vC}
	\fmf{plain}{v6,vC}
	\fmf{plain}{v7,vD}
	\fmf{plain}{v8,vD}
	\fmf{plain}{vA,vX}
	\fmf{plain}{vB,vX}
	\fmf{plain}{vC,vY}
	\fmf{plain}{vD,vY}
	\fmf{plain}{vX,vY}
\end{fmfgraph*}
}
\end{fmffile}
\ .
\end{equation*}
In a channel where an external operator meets two internal operators, we cannot set both internal operators to the identity, since this would vanish.  In this case, we contract into the identity operator whenever possible, and set other internal weights to $h$. This is what we mean by the `vacuum block' $f_0$ in the general case.  It can always be computed by imposing trivial monodromies on the appropriate cycles.  This is illustrated by the 6-point example above, where the vacuum block has weights
\vspace{.5cm}
\begin{equation}\label{ublockex}\small
\begin{fmffile}{sixB}
\begin{fmfgraph*}(50,15)
  	\fmfleft{i1,i2}
        \fmfright{o1,o2}
	\fmftop{t1,t2,t3,t4,t5,t6}
	\fmfbottom{b1,b2,b3,b4,b5,b6}
        \fmf{plain}{i1,v1,i2}
        \fmf{plain}{o1,v2,o2}
        \fmf{plain,label=0}{v1,vA}
	\fmf{plain,label=$h$}{vA,vB}
	\fmf{plain,label=0}{vB,v2}
	\fmf{plain,tension=0}{t3,vA}
	\fmf{plain,tension=0}{b4,vB}
	\fmflabel{1}{i1}
	\fmflabel{2}{i2}
	\fmflabel{3}{t3}
	\fmflabel{4}{o2}
	\fmflabel{5}{o1}
	\fmflabel{6}{b4}	
\end{fmfgraph*}
\end{fmffile}
\end{equation}
This is computed by imposing trivial monodromy $M = 1_{2x2}$ on the cycles $\gamma_{p}$ and $\gamma_r$ in (\ref{cycb}) (which automatically fixes the monodromy on $\gamma_q$ to correspond to $h$).

This procedure can be implemented numerically, though we will not need the explicit results.  
 
In the rest of this subsection we will sketch a derivation of these statements, closely following the logic  used for the 4-point function  \cite{zam1,Hadasz:2005gk,Harlow:2011ny}.  The idea is to consider the Liouville correlation function $\langle \chi O_1  \cdots O_k\rangle$, where $\chi$ is a light operator corresponding to a null state, $\langle \chi | \chi \rangle = 0$.  The null decoupling equation for this correlator is the differential equation (\ref{diff1}), and the monodromy condition projects onto the contribution from a particular conformal family.

In Liouville theory (see \cite{Zamolodchikov:1995aa,Harlow:2011ny,Hadasz:2005gk}
for reviews that we follow closely), it is convenient to parameterize the central charge as
\be
c = 1 + 6 Q^2 \ , \quad Q = b + \frac{1}{b} \ .
\ee
Local operators $V_\alpha = e^{2\alpha \phi}$ have dimension $\Delta_\alpha = \alpha(Q-\alpha)$. Correlation functions are defined by the path integral
\be
G_k(z_i)\equiv \langle V_{\alpha_1}(z_1) \cdots V_{\alpha_k}(z_k) \rangle = \int D\phi\, V_{\alpha_1}(z_1) \cdots V_{\alpha_k}(z_k) e^{-Q^2 S_L[\phi]}
\ee
with the Liouville action 
\be
S_L[\phi] = \frac{1}{4\pi}\int d^2 z\left(|\p\phi|^2 )+ \mu e^{\phi}\right) + \mbox{boundary terms} \ .
\ee
In the semiclassical limit $b\to 0$, correlation functions of heavy operators are given by the action of a classical saddlepoint,
\be\label{gkup}
G_k(z_i) \approx e^{-\frac{c}{6}S_{cl}} \ ,
\ee
where the saddlepoint obeys boundary conditions at the insertion points that depend on $h_i/c$. 

Let us choose an OPE channel, and denote the contribution from a particular choice of internal weights by $G_k(z_i)|_{p,q,\dots}$. That is,
\be\label{gkope}
G_k(z_i)|_{p,q,\dots} = c_p c_q \cdots \bF(z_i)\bar{\bF}(\bz_i) \ ,
\ee
where $c_a$ is the OPE coefficient computed exactly by DOZZ \cite{Dorn:1994xn,Zamolodchikov:1995aa}. (The other two indices on $c_a$ are suppressed, and depend on the channel.) In the semiclassical limit, $G_k$ on the l.h.s. and the $c_a$ on the r.h.s. can all be represented as $e^{-\frac{c}{6}S_{cl}}$, so the blocks $\bF$ have the exponential form (\ref{blockup}).

The monodromy prescription for the semiclassical block comes from studying the decoupling equation for a light null state.  The light operator $\hat{\psi}$ with $\Delta_\psi = -\frac{1}{2} - \frac{3}{4}b^2$ has a null state at level two,
\be\label{nulls}
|\chi\rangle = \left(L_{-2}  - \frac{3}{2(2\Delta_\psi + 1)}L_{-1}^2\right)|\psi\rangle \ , \quad\quad \langle \chi | \chi \rangle = 0 \ .
\ee
If we insert this operator into the $k$-point correlator,
\be\label{defpsi}
\Psi = \langle \hat\psi(z) V_{\alpha_1}(z_1)\cdots V_{\alpha_k}(z_k)\rangle \ ,
\ee
then decoupling of the null state implies \cite{Belavin:1984vu}
\be\label{dec}
\left[\p_z^2 - \frac{2(2\Delta_\psi + 1)}{3}\sum_{i=1}^k\left(\frac{h_i}{(z-z_i)^2} + \frac{\p_{z_i}}{(z-z_i)}\right)\right]\Psi = 0 \ .
\ee
Expanding the heavy operators in conformal blocks, denote the contribution of a particular set of primaries by $\Psi_{p,q,\dots}$.  The different $\Psi_{p,q,\dots}$ have different monodromies as we move $z$ around the $z_i$, so (\ref{dec}) must be true for each contribution individually.

Now we will evaluate $\Psi_{p,q,\dots}$ semiclassically to simplify (\ref{dec}).  The insertion of a light operator does not change the classical saddlepoint, so the path integral gives
\be
\Psi_{p,q,\dots} \approx \psi(z)c_pc_q\dots \exp\left(-\frac{c}{6}(f(z_i) + \bar{f}(\bz_i))\right) , 
\ee
where $\psi$ is a classical field that can be interpreted as the expectation value of $\hat{\psi}$ in the presence of the heavy operators. Plugging into the decoupling equation (\ref{dec}), it becomes the differential equation (\ref{diff1},\ref{diff2}) with accessory parameters $c_i = \p f /\p z_i$ as claimed.

The last step is to relate the monodromies of $\psi(z)$ to the weights $h_a$ of the operators running in the OPE. Suppose the channel involves a contraction $O_aO_b \to O_c$. Then if $\psi(z)$ is inserted somewhere on a cycle enclosing $O_a$ and $O_b$ that is taken small enough to avoid other operators, when we do the OPE to produce the conformal block expansion of (\ref{defpsi}) we come across $\langle O_a(z_a) O_b(z_b) \psi(z) O_c(z_c)\rangle$. This correlator satisfies a null decoupling equation like (\ref{dec}). Plugging into this decoupling equation the leading OPE behavior $\psi(z)O_c(z_c) \sim (z-z_c)^\kappa O_d(z_c)$ and looking around $z \sim z_c$ fixes
\be
\kappa = \half (1 \pm \Lambda_c) \ , \quad \mbox{where} \quad h_c = \frac{c}{24}(1-\Lambda_c^2) \ .
\ee
Finally, the monodromy of the two solutions $(z-z_c)^{\half(1 \pm \Lambda_c)}$ on a cycle enclosing $z_c$ is
\be
M_c = -\begin{pmatrix}e^{-i\pi\Lambda_c} & 0\\ 0 & e^{i\pi \Lambda_c}\end{pmatrix} \ , \quad\quad \Tr M_{c} = -2\cos \pi \Lambda_c \ .
\ee
This is the same as the monodromy around $z_a$ and $z_b$, so this completes the argument.

\subsection{Entanglement entropy}
Given the monodromy prescription for $k$-point conformal blocks, the discussion of the entanglement entropy for $N$ intervals is essentially identical to that for two intervals in section \ref{s:twointervals}. Region $A$ is taken as in (\ref{amany}).  The twist correlator (\ref{gent}) is expanded in conformal blocks in a particular channel with $k=2N$ external operators.

The leading contribution in this channel comes from light operators exchanged between the coincident $\Phi_+$ and $\Phi_-$.  We can replace these operators with the identity, because the semiclassical block depends only on $h_p / c \to 0$ and the overall coefficient will be irrelevant when we take the log. Therefore the leading non-vanishing contribution is what we defined as the vacuum block in the previous subsection,
\be\label{multr}
S_A^{(n)} \approx \frac{nc}{3(n-1)}f_0\left(\frac{H}{nc},z_i\right) \ .
\ee
Because the blocks depend exponentially on $c$, in the large-$c$ limit this is the full answer to all orders in a series expansion in this channel.  

The conclusion is that for any channel, the Renyi entropy is given by (\ref{multr}) in a finite range of $z_i$, with $f_0$ defined by imposing trivial monodromies in a way dictated by the channel.  As with $N=2$, we have not ruled out the possibility that there are ranges of $z_i$ where the dominant contribution is not the vacuum block in any channel. The numerical calculation of the Renyi entropies from (\ref{multr}) is straightforward but computationally intensive, since we do not have the advantage of a fast recursion formula as we did for the four-point blocks.

The derivation of the entanglement entropy proceeds as in section \ref{s:twointervals}.  We must impose trivial monodromies on certain cycles of the differential equation (\ref{diff1},\ref{diff2}) with external weights $h_i = H$ defined in (\ref{tdim}) and take $n\to 1$. For $n\to 1, H \to 0$, all of the terms in $T(z)$ are important only at the singular points, so the problem decouples into $N$ independent monodromy problems.  In each case, if we pair $\Phi_+(z_i)$ with $\Phi_-(z_j)$, then setting $n=1+12\alpha$ we find $c_i = 12\alpha/(z_i - z_j) + O(\alpha^2)$. Therefore the semiclassical vacuum block is
\be\label{fmult}
f_0(\alpha,z_i) = 12\alpha \sum_{(i,j)}\log (z_i - z_j) + O(\alpha^2)\ ,
\ee
and the entanglement entropy is (restoring the UV cutoff)
\be\label{finals}
S_A = \frac{c}{3} \sum_{(i,j)} \log \left(z_i - z_j\over \epsilon\right) \ .
\ee
The sum is over pairs $(i,j)$ dictated by the OPE channel, selected as follows.  First, we pair operators which are directly contracted in the OPE; in the 6-point example, these are $(1,2)$ and $(4,5)$.  Then, we pair operators connected in the OPE tree diagram by a string of internal $H$'s, like the pair $(3,6)$ in (\ref{ublockex}).  These are the correct pairings because these are the ways we can divide the OPE tree diagram cutting only identity operators, so these correspond to the cycles of trivial monodromy.

The result (\ref{finals}) is in a particular channel. Each channel is an upper bound for the full answer.  If we assume that the regions where the vacuum block dominates in various channels cover $z_i$ everywhere on the real line, then the full answer is obtaining by selecting the channel where this is a minimum,
\be\label{minans}
S_A = \mbox{min} \ \frac{c}{3}\sum_{(i,j)}\log\left(z_i - z_j \over \epsilon\right) \ .
\ee
\begin{figure}
\begin{center}
\includegraphics{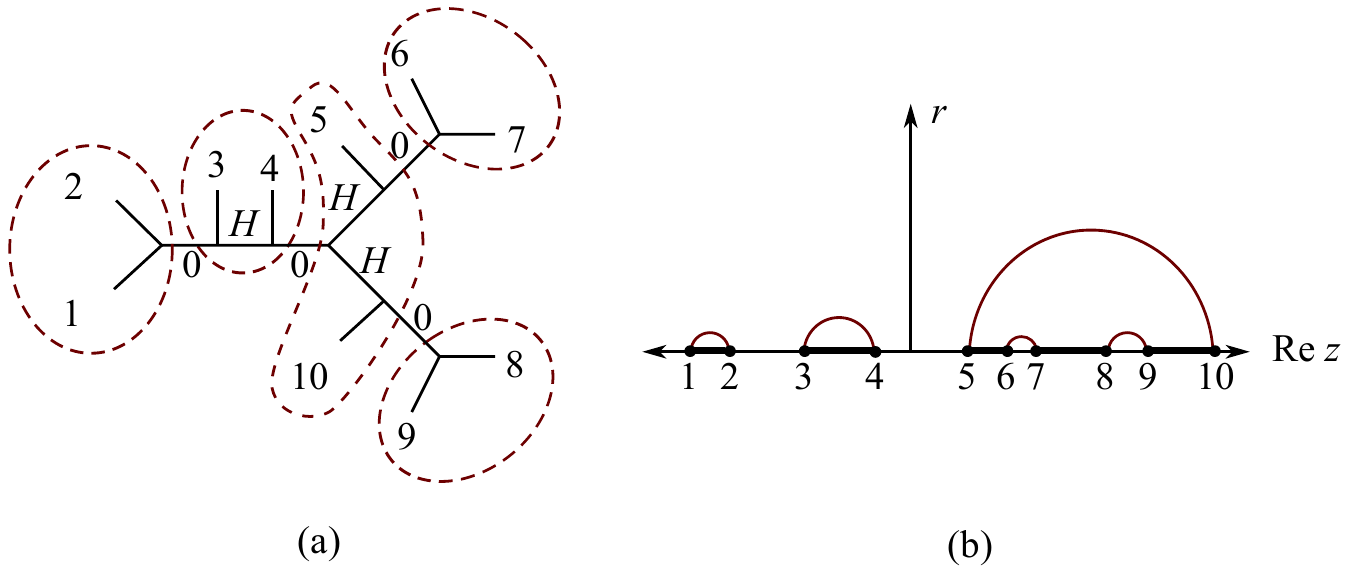}
\end{center}
\caption{Example of the entanglement entropy of 5 disjoint intervals at large central charge in a particular channel, (a) in CFT and (b) holographically. In (a), trivial monodromy is imposed on cycles in the $z$-plane corresponding to the dashed lines.  The solid semicircles in (b) are geodesics in AdS$_3$ with radial direction $r$, and $A$ is the shaded region on the boundary. \label{fig:complicated}}
\end{figure}

The result (\ref{minans}) is identical to the holographic formula for entanglement entropy \cite{Ryu:2006bv}.  Each term is $\frac{1}{4G_N} = \frac{c}{6}$ times the length of a geodesic through a constant-time slice of AdS$_3$, connecting the two twist operators. 

A more complicated example is illustrated in figure \ref{fig:complicated}, in bulk and boundary.  In the CFT, we draw the OPE tree diagram, and place a unit operator `$0$' or twist operator `$H$' on each internal line. Then we pair the twist operators by drawing cycles which are only allowed to pierce internal lines carrying a unit operator. The semiclassical vacuum block $f_0$ is computed by imposing trivial monodromy on the corresponding cycles in the $z$-plane. Finally, the analytic continuation to $n=1$ equals the entanglement entropy computed by the length of geodesics in AdS$_3$ connecting the paired operators. As a consistency check, we confirmed numerically that accessory parameters obtained by differentiating (\ref{fmult}) indeed lead to trivial monodromies in this 10-point example.

\section{Discussion}

The universal behavior of entanglement entropy in 2d CFTs with holographic duals is in some ways an extension of Cardy's formula for the thermodynamic entropy \cite{Cardy:1986ie}. The Cardy formula holds universally at high temperature $T \to \infty$, but in theories with a holographic dual, it applies above a critical temperature $T_c = 1$. The sharp transition at $T_c$ is a consequence of the small number of light operators.  It is holographically dual to the Hawking-Page phase transition between black holes and a thermal gas in AdS$_3$, and the Cardy formula equals the black hole entropy \cite{Witten:1998zw,Strominger:1997eq,Maldacena:1998bw}. This gives a statistical origin of black hole entropy in many examples.

According to the Ryu-Takayanagi formula, entanglement entropy is a way of generalizing this success to geometric surfaces other than black hole horizons.  Patching together spacetime via entanglement may be a useful way to approach quantum gravity \cite{VanRaamsdonk:2010pw}, and if so, these minimal surfaces must play an important role.
 The universal formula for the entanglement entropy of a 2d CFT provides the microscopic origin for the area of these surfaces.  Like the Cardy formula, it is really only semi-microsopic, in the sense that it relies on a quantum field theory with a microscopic definition but does not involve the detailed description of states in that theory.

The gravitational sector of the AdS/CFT correspondence in the semiclassical limit can be summarized by the statement that gravity in $d+1$ dimensions is the theory of stress tensors in $d$-dimensional CFT.  That is, gravity describes the thermodynamic sector of the CFT. In $d=2$, the stress tensor lives in the same Virasoro representation as the vacuum state, so with no other matter present this becomes the statement that 3d gravity at weak coupling is the theory of the Virasoro vacuum representation at large $c$.  On trivial topology this is the fact that gravitons are Virasoro descendants \cite{Brown:1986nw}; on a torus it implies the Cardy formula for black hole entropy; and one consequence at higher genus is (modulo the caveats in the derivation) the holographic formula for entanglement entropy.

Semiclassical Liouville theory provides a link between these two descriptions, gravity and CFT.  On the one hand, the on-shell action of 3d gravity is equal to a Liouville action \cite{Verlinde:1989ua,Coussaert:1995zp,Krasnov:2000zq}.  On the other hand, since the thermodynamics of these CFTs is universally determined by the algebra, it can be computed in Liouville theory at large $c$. Note that this does not mean quantum Liouville theory is equivalent to either gravity or the CFT (see \cite{Witten:2007kt,Maloney:2007ud} for obstacles to this interpretation). Rather, Liouville is useful because in the semiclassical limit it naturally computes thermodynamic quantities on both sides \cite{Martinec:1998wm}. 

To be more specific, a natural conjecture extending our results is that in the class of theories we have considered, the leading-$c$ partition function on any Riemann surface $M_2$ is
\be
\log Z_{CFT} = -\frac{c}{3} \min_{\Gamma} S_{ZT}(M_2, \Gamma) \ ,
\ee
where $\Gamma$ indicates the choice of cycles necessary to specify the Zograf-Takhtajan Liouville action discussed briefly in section \ref{ss:compare}. See \cite{Headrick:2012fk} for related comments about universality in these theories. It would be very interesting to prove this statement, and to understand exactly what restrictions must be placed on the class of CFTs to do so.

This discussion is of course special to two-dimensional CFT.  In $d>2$, the vacuum representation is trivial.  The analogous calculation would be to include the exchange of all operators built from the stress tensor and its derivatives.  This is a much harder problem, because there is no local $d$-dimensional field theory that captures this sector of the CFT, like Liouville does in two dimensions.  On the other hand, in a 3d CFT with higher spin symmetry, correlation functions of currents are fixed by symmetry \cite{Maldacena:2011jn}. This is similar to the situation in 2d CFT, so it is plausible that a Liouville-like classical theory connects these theories to their higher-spin gravity duals (see \cite{Giombi:2012ms} and references therein).

\bigskip

\noindent{\bf Supplemental material} \nopagebreak

\noindent
A \textit{Mathematica} notebook implementing the conformal block recursion, the numerical calculation of monodromies and the accessory parameter $c_2$, and the consistency check of the 10pt example in figure \ref{fig:complicated} is provided with the arXiv submission for download.

\bigskip

\noindent{\bf Acknowledgments} \nopagebreak

\noindent
It is a pleasure to thank M.~Cheng, T.~Faulkner, M.~Headrick, C.~Keller, and J.~Maldacena for essential discussions. This work was supported in part by U.S.~Department of Energy grant DE-FG02-90ER40542 and by the Corning Glass Works Foundation Fellowship Fund.

\bigskip

\appendix
\section{Calculations of the conformal block}\label{app:numerics}
In this appendix we discuss the calculation of the 4-point Virasoro block $\bF$ and the semiclassical block $f$, and compute the Renyi entropy for two intervals in a series expansion. A \textit{Mathematica} notebook implementing the recursion relation and the numerical monodromy algorithm is provided with the arXiv submission of this paper.

The Virasoro block may be computed by a brute-force series expansion, but this tends to be the slowest method for numerical calculations.  Two efficient recursion formulae exist, one using the analytic structure as a function of $c$ \cite{Zamolodchikov:1985ie} and the other as a function of the internal weight $h_p$ \cite{zam1} (see also \cite{Zamolodchikov:1995aa}). The latter converges very quickly for $h_i \sim c$, so the first few terms can be used to quickly compute $\mathcal{F}$ for, say, $|x| \lesssim 0.99$. It is an expansion in
\be
q(x) = e^{-\pi K(1-x)/K(x)}
\ee
where $K$ is the complete elliptic integral of the first kind.  The inverse is
\be
x = \left(\frac{\theta_2(\tau)}{\theta_3(\tau)}\right)^4 \ , \quad q\equiv e^{i\pi \tau} \ .
\ee
(Note that this is the same as the transformation to the torus in \cite{Lunin:2000yv,Headrick:2010zt}). The recursive formula for the 4-point conformal block on the sphere is:
\begin{align}\label{zamrec}
\bF(c, h_p, h_i, x) &= (16q)^{h_p + \frac{1-c}{24}}x^{\frac{c-1}{24} - h_1 - h_2}(1-x)^{\frac{c-1}{24} - h_2 - h_3}\theta_3(q)^{\frac{c-1}{2} - 4 \sum h_i}H(c,h_p, h_i, q)\notag\\
H(c,h_p, h_i, q) &= 1-\half \sum_{m,n>0} \frac{(16q)^{m n}A_{mn}}{B_{mn}(h_p - H_{mn})}H(c, H_{mn} + mn, h_i, q)\notag\\
\alpha_\pm(c) &= \sqrt{1-c\over 24} \pm \sqrt{25-c\over 24}\\
 H_{mn}(c) &= \frac{c-1}{24} +S_{mn}^2\notag\\
A_{mn}(c,h_i) &=  \prod_{p}\prod_{q} (\lambda_3 + \lambda_4 -  S_{pq})(\lambda_3 - \lambda_4 -  S_{pq})\notag\\
& \quad\quad\quad\quad\quad\quad \quad\quad\quad\times (\lambda_1 + \lambda_2 -  S_{pq})(\lambda_2-\lambda_1 -  S_{pq})\notag\\
B_{mn}(c) &= \prod_{ a=-m+1}^m  \prod_{b=-n+1}^n 2 S_{ab}\notag\hspace{7cm}\\
\lambda_i(h_i,c) &= \sqrt{h_i + \frac{1-c}{24}}\notag\\
S_{rs}(c) &= \half(\alpha_+ r + \alpha_- s)\notag
\end{align}
Zeroes are omitted in the product $B_{mn}$. The product in $A_{mn}$ is taken over $p=-m+1, -m+3, \dots, m-1$ and $q=-n+1, -n+3, \dots, n-1$. For $h_p = 0$ with equal external weights, $(m,n) = (1,1)$ is omitted in the sum on the second line. For analytic calculations, it is useful to parameterize $c = 1 + 6 (b+ 1/b)^2$.

The monodromy problem described in section \ref{genop} can also be solved numerically to compute the semiclassical block $f$ or $f_0$.  This is done by computing the monodromy $\Tr M_{0x}$ at fixed accessory parameter $c_2$, solving for $c_2$ by shooting, then integrating with the boundary condition (\ref{fseries}) to obtain $f$. For a fixed value of the accessory parameter $c_2$, the monodromy invariant of the differential equation can be computed by numerically solving the ODE on a circle in the $z$-plane with two arbitrary initial conditions to find $\psi_{1,2}$. Equation (\ref{monoto}) and its derivative fix the entries of $M$.

Now we will use the recursion formula to compute the Renyi entropy of two intervals in a series expansion around $x\sim 0$ at large $c$. With central charge $c_0$, equal external weights $h_i = c_0 \delta$ and internal weight $h_p = 0$, from (\ref{zamrec}) we find $\bF \approx e^{-\frac{c_0}{6}f_0}$ with
\begin{align}
f_0(\delta,x) &= 12 \delta \log x +a_2 q^2 + a_4 q^4 + a_6 q^6 + O(q^8)\\
a_2 &= -3072 \delta^2\notag\\
a_4 &=  \frac{24576}{5}\delta^2(1-64\delta+704\delta^2) \notag\\
a_6 &= \frac{4096}{35}\delta^2(195 - 18432 \delta + 784384\delta^2 - 14548992\delta^3 + 98566144\delta^4)\notag
\end{align}
The Renyi entropy is
\be\label{appr}
S_A^{(n)}= \frac{cn}{3(n-1)}f_0\left(\frac{1}{24}(1-\frac{1}{n^2}), x\right) \ .
\ee
For comparison to \cite{Headrick:2010zt}, the Renyi mutual information is defined by subtracting the log term from (\ref{appr}).  The result is equation (4.41) of \cite{Headrick:2010zt}; when $n=2$ it agrees with equation (4.30) of \cite{Headrick:2010zt}.  

\section{Phases of the $N=n=2$ Renyi entropy}\label{app:torus}

In this appendix we argue that for $N=n=2$, the only phase transition in the Renyi entroy is at $x = \half$, using crossing symmetry as well as some properties of $f$ which are unproven but checked numerically.  (This was derived by a different method in \cite{Headrick:2010zt}.) Thus the Renyi entropy $S_A^{(2)}$ is always dominated by the vacuum block in one channel. For this calculation we will also assume that the OPE coefficients of light operators with $
\Delta < \Delta_{gap}$ do not grow exponentially with $c$.  

The goal is to show that the contribution of heavy operators in the $t$-channel is exactly equal, to leading order in $1/c$, to the vacuum contribution in the $s$-channel, and vice-versa.  First we observe that the sum over light states is always dominated by the vacuum state at any value of $x \leq \half$, not just near $x=0$:
\be\label{lightsum}
\sum_{p|\delta_p < 1/24} a_p e^{-\frac{nc}{3}f(\delta_p, \beta_n, x)} \sim e^{-\frac{nc}{3}f_0(\beta_n,x)}  \quad \quad 0 < x <\half \ ,
\ee
where
\be
\beta_n  = \frac{H}{nc} \ , \quad\quad \delta_p = \frac{h_p}{nc} \ .
\ee
This follows from the fact that for $\beta_n \leq \frac{1}{24}$ and any fixed $x\leq \half$, $f(\delta_p, \beta_n, x)$ is an increasing function of $\delta_p$ in the range $0 < \delta_p < 1/24$. This is straightforward to check numerically, and holds for any $n$.

Now, crossing symmetry is the statement that the expansions in the two channels are equal:
\be\label{brak}
\sum_p a_p \left[e^{-\frac{nc}{3}f(\delta_p,\beta_n,x)} - e^{-\frac{nc}{3}f(\delta_p,\beta_n,1-x)}\right] = 0 \ .
\ee
For each term in the sum, we want to keep only the dominant exponential.  For this we need the behavior of the function
\be\label{mono}
D(\delta,\beta,x) \equiv f(\delta, \beta, 1-x) - f(\delta,\beta,x) \ .
\ee
When $D$ is positive, the $s$-channel term in (\ref{brak}) dominates, and when it is negative, the $t$-channel term dominates.  Numerically, we find the following: Depending on $\delta,\beta$, there are three different behaviors for $D$ as a function of $x$.  $D$ is monotonically increasing as a function of $x$ for large $\delta$, monotonically decreasing for small $\delta$, and non-monotonic (with multiple roots) in some intermediate range.  When $n < 2$, the intermediate range of $\delta$ lies below $\delta = 1/24$, and when $n >2$ it lies above $\delta = 1/24$. For $n=2$, the intermediate range disappears; $D$ is decreasing for $\delta < 1/24$ and increasing for $\delta > 1/24$. We caution that these statements rely on a numerical calculation of the conformal block near the region where the series diverges, $x \to 1$.  This means that we have actually checked them only on the interval $x \in [\epsilon, 1-\epsilon]$ with $\epsilon \sim 0.01$.  We will assume that the statements hold exactly, but the numerical evidence is not sufficient to be entirely convincing.

Returning to (\ref{brak}), if we take $x < \half$ and $n \leq 2$, then these properties of $D$ imply that the crossing equation becomes
\be
e^{-\frac{nc}{c}f_0(\delta_p, \beta_n, x)}  - \sum_{p|\delta_p>1/24} a_p e^{-\frac{nc}{3}f(\delta_p, \beta_n, 1-x)} = 0 \ ,
\ee
where we have dropped an irrelevant prefactor on the first term.  This is exactly what we set out to show: the vacuum contribution in the $s$-channel is equal to the contribution of all heavy operators in the $t$-channel. Although this argument formally holds for $1<n \leq 2$, it tells us nothing about the entanglement entropy $n=1$ because we should not trust the analytic continuation from the single integer point $n=2$.

This argument fails for $n > 2$ because $D$ is not monotonic in the range $\delta > 1/24$, so the dominant contributions to (\ref{brak}) do not separate cleanly into light operators in the $s$-channel and heavy operators in the $t$-channel. In fact, using the Cardy growth of states above $\delta = 1/24$, it can be shown that to avoid other phases for $n>2$, $a_p$ must be exponentially suppressed in $c$ around $\delta \sim 1/24$.  It would be interesting to explore whether $a_p$ can be bounded some other way to demonstrate this suppression, perhaps along the lines of \cite{Pappadopulo:2012jk}.

\begin{spacing}{1.1}

\end{spacing}

\end{document}